# Thermal Transport Across Graphene Step Junctions

Miguel Muñoz Rojo[1,†], Zuanyi Li[1,†], Charles Sievers[2], Alex C. Bornstein[1], Eilam Yalon[1], Sanchit Deshmukh[1], Sam Vaziri[1], Myung-Ho Bae[3], Feng Xiong[4], Davide Donadio[2] and Eric Pop[1,5,6,*]

[1]*Department of Electrical Engineering, Stanford University, Stanford, CA 94305, USA*
[2]*Department of Chemistry, University of California Davis, Davis, CA 95616, USA*
[3]*Korea Research Institute of Standards and Science, Daejeon 34113, Republic of Korea*
[4]*Department of Electrical & Computer Engineering, University of Pittsburgh, Pittsburgh, PA 15261, USA*
[5]*Department of Materials Science & Engineering, Stanford University, Stanford, CA 94305, USA*
[6]*Precourt Institute for Energy, Stanford University, Stanford, CA 94305, USA*
[*]*Corresponding authors e-mail: epop@stanford.edu and ddonadio@ucdavis.edu*
[†]*These authors contributed to this work equally*

**Abstract:**

Step junctions are often present in layered materials, i.e. where single-layer regions meet multi-layer regions, yet their effect on thermal transport is not understood to date. Here, we measure heat flow across graphene junctions (GJs) from monolayer to bilayer graphene, as well as bilayer to four-layer graphene for the first time, in both heat flow directions. The thermal conductance of the monolayer-bilayer GJ device ranges from ~0.5 to $9.1 \times 10^8$ Wm$^{-2}$K$^{-1}$ between 50 K to 300 K. Atomistic simulations of such GJ device reveal that graphene layers are relatively decoupled, and the low thermal conductance of the device is determined by the resistance between the two distinct graphene layers. In these conditions the junction plays a negligible effect. To prove that the decoupling between layers controls thermal transport in the junction, the heat flow in both directions was measured, showing no evidence of thermal asymmetry or rectification (within experimental error bars). For large-area graphene applications, this signifies that small bilayer (or multilayer) islands have little or no contribution to overall thermal transport.

1. Introduction

The emergence of two-dimensional (2D) materials has brought new opportunities to explore fundamental physical properties and to exploit these materials for new applications [1]. As the first isolated 2D material [2–4] and due to its extraordinary transport properties [5], graphene has been extensively studied especially for electronic applications. However, the properties of graphene can be altered due to crystal imperfections which appear, for example, during graphene growth by chemical vapor deposition (CVD). One such type are grain boundaries (GBs) [6], i.e. line defects where two graphene grains (of the same thickness) are stitched together. Other defects are graphene junctions (GJs), i.e. the steps between regions with different number of graphene layers, such as monolayer-to-bilayer (1L-2L) junctions.

The properties of GBs are relatively well understood, having been measured electrically [7], thermally [8], or mechanically [9]. For example, GBs reduce the overall electrical [7] and thermal conductivity [10,11] of graphene due to electron and phonon scattering, respectively. However, GJs have only recently attracted more interest with few experimental studies of their properties in electronics [12], optoelectronics [13], and as p-n junctions [14]. A theoretical study assigned thermal rectification properties to GJs [15], however this has not been examined experimentally. Other simulations also showed that heat transfer at GJs is non-trivial, because in the multilayer region different layers may have different temperatures [16]. Knowledge of heat flow across GJs is important not just fundamentally, but also for practical applications in terms of how they modify the overall thermal conductivity of graphene (as GBs do [10,17]), or where GJs could act as phonon filters). As an example, electronic devices based on graphene and other 2D materials often contain GJs, but little is known about how their thermal resistance affects the overall device performance [18,19].

Here, we investigate for the first time the temperature-dependent heat flow across GJs supported on $SiO_2$ substrates. Our experimental results combined with molecular and lattice dynamics simulations indicate thermal decoupling between layers caused by a large thermal boundary resistance (TBR). Thus, we establish a microscopic understanding of thermal conduction across GJs and clarify their role in large-area thermal management applications of graphene.

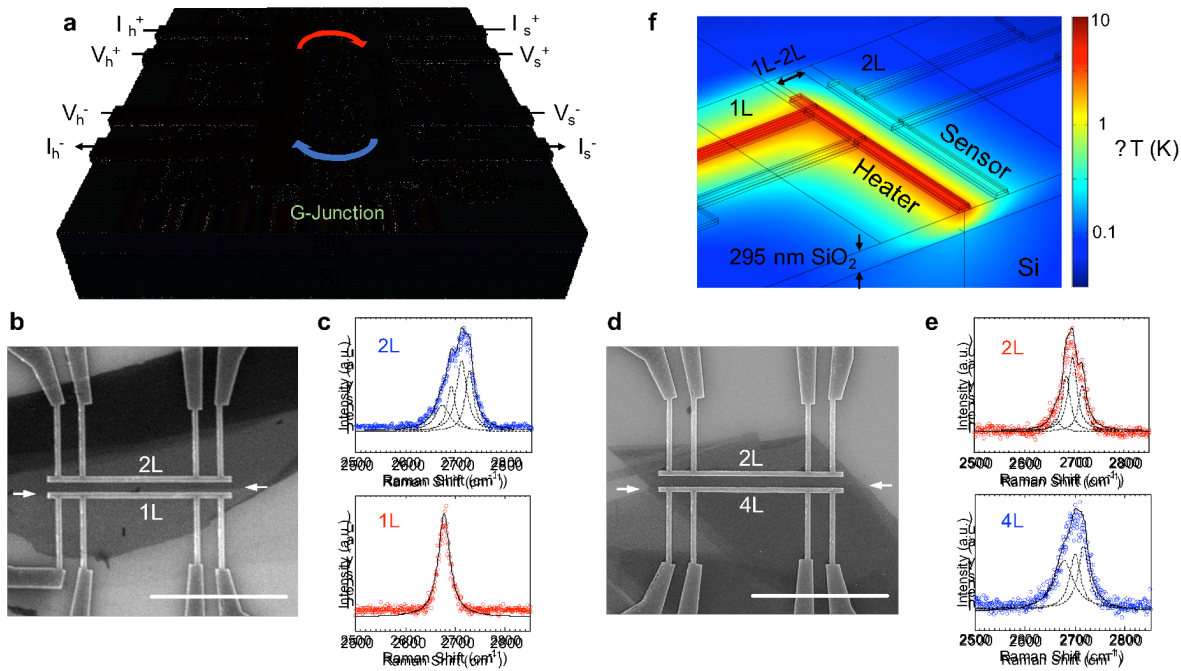

**Figure 1.** (a) Device layout for thermal conductance measurements across 1L-2L graphene junction (GJ). Two metal lines with ~400 nm separation were formed with the GJ between them. A thin $SiO_2$ layer under the metal lines provides electrical isolation and thermal contact with the graphene beneath. One of the lines is used as heater while the other one as sensor. The heater and sensor can be reversed to measure the heat flow in both directions. (b,d) and (c,e) show scanning electron microscopy (SEM) images and Raman spectra of the 1L-2L and 2L-4L graphene junctions, respectively. GJs are indicated by arrows and all scale bars are 5 μm. The dashed lines represent Lorentzian fits of the Raman spectra. (f) Three-dimensional (3D) simulation of the experimental structure, showing temperature distribution with current applied through the heater.

2. Experimental results

Figure 1a illustrates the schematic of the device structure we used to measure the thermal conductance across GJs. Graphene used in this study (see Methods) is mechanically exfoliated onto a $SiO_2$/Si substrate (Supplementary Section 1) and junctions were identified by optical microscopy, atomic force microscopy (AFM), Raman spectroscopy (see Methods) and were finally confirmed by scanning electron microscopy (SEM) after all measurements were completed. During thermal measurements the two lines were used as the heater and thermometer, interchangeably [20,21]. A thin layer of $SiO_2$ (~40 nm, electron-beam evaporated, see Methods) underneath the metal lines provided electrical isolation from the graphene. Figures 1b and 1d show SEM images of the two devices measured, which correspond to 1L-2L and 2L-4L (bilayer to four-layer) GJs, respectively. Figures 1c and 1e show Raman spectra obtained on each side of the GJ, de-termining the number of graphene layers. The Raman spectra do not show discernible D peaks even after patterning the metal lines, confirming relatively defect-free, crystalline graphene regions (Supplementary Section 2).

We performed heat flow measurements from 50 K to 300 K on these GJ samples and on similar control samples without graphene. We also measured heat flow across the GJs in both directions by swapping the heater and sensor, to test for possible asymmetry in the heat flow as a consequence of phonon scattering at the junction, which would lead to thermal rectification for large temperature differentials [15]. The measurements are performed as follows. Current is forced into a metal line, which acts as a heater, while both metal lines are used to sense temperature, setting up a temperature gradient across the GJ. The metal lines are thermo-resistive elements, which allow us to convert measured changes of electrical resistance into variation of the temperature of the sensor, $\Delta T_S$, and heater, $\Delta T_H$, as a function of the heater power $P_H$ (Supplementary Section 4). We calibrated both metal lines for each sample by monitoring the resistance over a slightly wider temperature range, from 40 K to 310 K, to determine the temperature coefficient of resistance (TCR) and quantify temperature variations (Supplementary Sections 6 and 7).

Once the temperature difference between the metal lines is known as a function of the heater power, the thermal conductance across the junction is obtained by processing the experimental data using a three-dimensional (3D) finite element model (FEM) [20,22] (see Methods). In this simulation, the graphene channel region between heater and sensor is treated with an effective thickness $h = 0.34n$ nm, where $n = 2$ in both devices because most of the two channels are covered by 2L graphene (see arrows in Figures 1b and 1d). In other words, the FEM fits the graphene channel with an effective thermal conductivity, $k$, between heater and sensor. The effective channel thermal conductance is $G = kh(W/L)$, where $W$ and $L$ are the graphene channel width and length.

The FEM shown in Figure 1f accurately replicates the experimental setup taking into account: i) all geometric dimensions of the metal lines, determined using SEM images (Supplementary Section 1); ii) the thickness of the $SiO_2$ under the graphene from ellipsometry (Supplementary Section 3) and its temperature-dependent thermal conductivity from measurements of the control sample (Supplementary Section 5); iii) the Si thermal conductivity for Si wafers with the same doping density [23] (Supplementary Section 3). The FEM also includes the effect of thermal boundary resistance (TBR) at Si-SiO$_2$ interfaces [20] from the control sample, graphene-SiO$_2$[24] and SiO$_2$-metal[25] interfaces, based on previous measurements of similar samples [20]. Figure 1f shows the simulated temperature distribution with current applied through the heater for the 1L-2L junction device. The thermal conductivity $k$ of the graphene channel is varied in the simulation until $\Delta T_S$ and $\Delta T_H$ vs. $P_H$ modeling results match well with the experimental data.

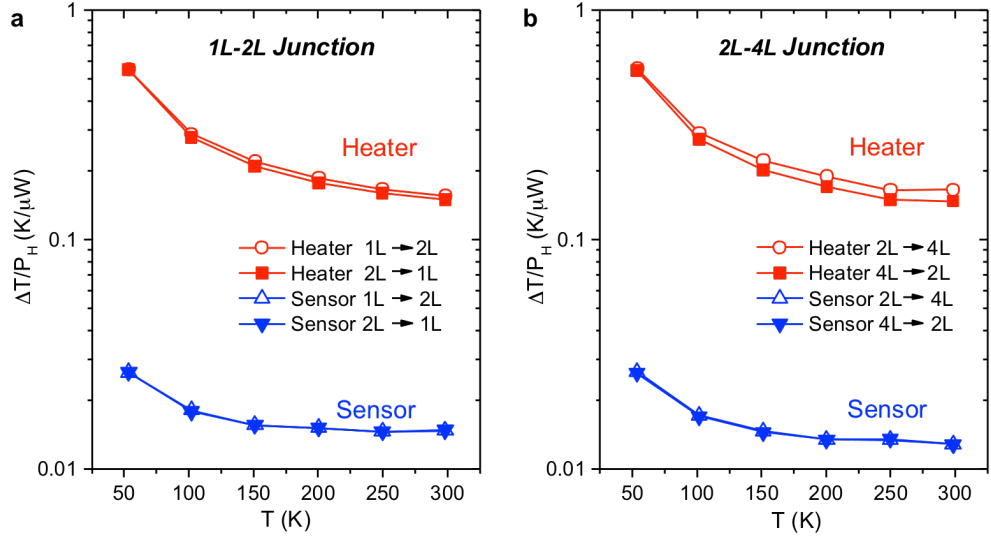

**Figure 2**. Experimental measurements of temperature rise in the heater and sensor per heater power, $\Delta T/P_H$, as a function of temperature for (a) the 1L-2L and (b) 2L-4L graphene junction. Heat flow was measured in both directions, from 1L → 2L vs. 2L → 1L, and from 2L → 4L vs. 4L → 2L, without observing thermal rectification. The uncertainty of these data is smaller than the symbol size.

We also measured a control sample without graphene in the channel to validate our method and to obtain the thermal properties of the parallel heat-flow path through the contacts, the supporting SiO$_2$, the SiO$_2$-Si interface and the Si substrate (Supplementary Section 5). These thermal properties obtained after processing the experimental data with the FEM show good agreement with well-known data from literature [20,26,27] over the full temperature range. Consequently, these data were used as inputs for the FEM simulation of the GJ structures.

Figure 2 shows the experimental heater temperature rise (in red) and sensor temperature rise (in blue) normalized by the heater power, $\Delta T/P_H$, as a function of temperature obtained for the two junctions studied, 1L-2L and 2L-4L. The heat flow was studied in both directions across the graphene junction to account for possible thermal rectification effects. The uncertainty of $\Delta T/P_H$ is ~0.5-1%, which agrees well with our previous experiments that use similar metal lines [20].

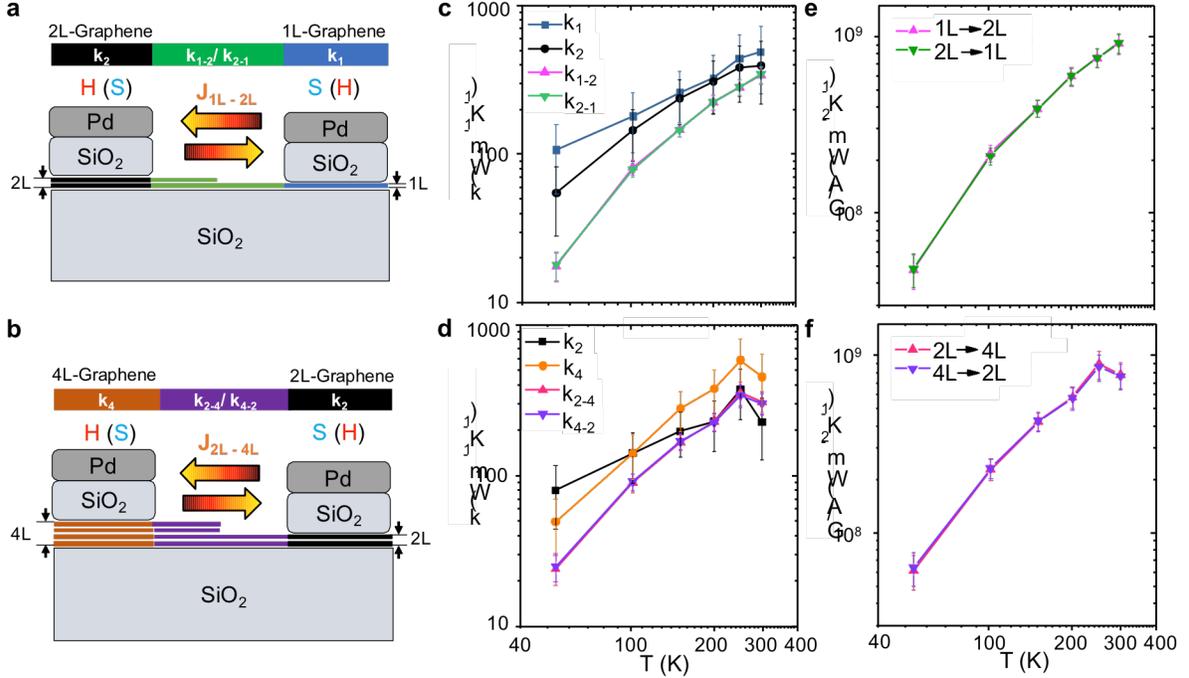

**Figure 3**. (a) and (b) Schematic cross-sections of 1L-2L and 2L-4L graphene junction experiments, respectively. The graphene layers and junctions are colored corresponding to different thermal conductivity regions, determined after processing the experimental data (Figure 2) with the FEM. (c) and (d) Thermal conductivity obtained for each region of graphene (1L, 2L and 4L) and for the 1L-2L and 2L-4L junctions for both heat flow directions. (e) and (f) Thermal conductance per unit area, i.e. thickness times width ($A = hW$), obtained at the junction. The results show no thermal rectification effect within the experimental error bars.

Figures 3a and 3b show the schematic of the 1L-2L and 2L-4L GJ samples. The rectangular colored sections on top illustrate the thermal conductivities, i.e. 1L ($k_1$), 2L ($k_2$) and 4L ($k_4$) for non-junction regions, while 1L-2L ($k_{1-2}$) and 2L-4L ($k_{2-4}$) represent the GJ channel region. These are used by the FEM to process the raw experimental data from Figure 2, yielding the effective thermal conductivities shown in Figures 3c-f. While the effective thermal conductivity of the GJ channel ($k_{1-2}$ and $k_{2-4}$) is determined from the temperature gradient between heater and sensor, the thermal conductivity of 1L, 2L and 4L is mainly determined from temperature variations only at the heater surroundings. Although not the main topic of this study, these supported 2L and 4L graphene thermal conductivity estimates are among the first of their kind (others being discussed below).

Figures 3c and 3d display the extracted thermal conductivity of the 1L, 2L and 4L graphene regions, as well as the effective thermal conductivity of the 1L-2L and 2L-4L junctions, in both directions of heat flow. The 1L, 2L and 4L thermal conductivities show similar values over the entire range of temperature. Their room temperature values are ~500 Wm$^{-1}$K$^{-1}$ for 1L, ~400 Wm$^{-1}$K$^{-1}$ for 2L, and ~450 Wm$^{-1}$K$^{-1}$ for 4L graphene, respectively. These are consistent with earlier measurements by Seol *et al.* [28], Sadeghi *et al.* [29], and by Jang *et al.* [21] who found the thermal conductivity of SiO$_2$-supported 1L, 2L and 4L graphene were ~580, ~600 and ~480 Wm$^{-1}$K$^{-1}$ at room temperature, respectively. To obtain the various thermal conductivities from the FEM fitting, we used the same TBR between graphene and SiO$_2$ for all layers, following Chen *et al.* [24], but there may be small differences in the TBR that could be behind this small variation. However, our results are in good agreement with values reported by Sadeghi *et al.* [29], which show that the thermal conductivity of SiO$_2$-supported graphene few-layers remains very similar.

In comparison, Figures 3c and 3d show that the effective thermal conductivity in the GJ regions, i.e. $k_{1-2}$ and $k_{2-4}$, is lower than in the graphene layers, i.e. $k_1$, $k_2$ and $k_4$. This difference becomes more evident as the temperature reduces from 300 K to 50 K. Figures 3e and 3f show the effective thermal conductance of the GJ regions, calculated by dividing the thermal conductivity with the metal line separation (see Methods). The thermal conductance for 1L-2L varies from 4.8 ± 1.1 × 10$^7$ Wm$^{-2}$K$^{-1}$ to 9.1 ± 1.2 × 10$^8$ Wm$^{-2}$K$^{-1}$ at 50 K and 300 K respectively, while for 2L-4L it varies from 6.1 ± 1.3 × 10$^7$ Wm$^{-2}$K$^{-1}$ to 7.7 ± 1.2 × 10$^8$ Wm$^{-2}$K$^{-1}$ at 50 K and 300 K respectively. Bae *et al.* [20] explained that as we shorten the length of a graphene channel, quasi-ballistic phonon transport effects reduce its thermal conductivity, because the longest phonon mean free paths become limited by the length of the channel. In other words, the graphene thermal conductivity is length-dependent in this sub-micron regime. The thermal conductivity of our GJ samples is consistent with values reported by Bae *et al.* [20] for length-dependent graphene without junctions. Additionally, that the thermal conductance of the 1L-2L and 2L-4L channels is almost identical for both heat flow directions, i.e. $k_{1-2} \approx k_{2-1}$ and $k_{2-4} \approx k_{4-2}$, indicates no measurable asymmetry in the heat flow or thermal rectification effects on supported graphene at the junction.

3. Molecular and Lattice dynamic simulations and discussion

To explain the measured thermal conductance of the GJs in both heat flow directions we consider two possible scenarios. The first scenario consists of thermal decoupling between the top and bottom layers of graphene, which could be attributed to the presence of a large TBR between layers. The thermal decoupling between layers would cause the heat to flow only through one layer, i.e. the bottom one, which would result in similar conductance values as the work of Bae *et al.* [20]. Moreover, the large TBR between layers would make phonon scattering at the junction negligible, which would support the idea of a non-asymmetry or thermal rectification effect. The second possible scenario would be a perfect coupling between the top and bottom graphene layers, i.e. very small TBR between layers, which would explain the similarity of the GJs thermal conductance with those shown by Bae *et al.* [20]. However, under these circumstances, we would expect the junction to scatter phonons more efficiently, which might induce some thermal asymmetry across the junction.

To quantitatively understand the phonon physics at the GJ, we performed atomistic molecular dynamics (MD) simulations and lattice dynamics (LD) calculations (see Methods). First, we evaluate a *suspended* 1L-2L junction by non-equilibrium molecular dynamics (NEMD) simulations [30] as shown in Figure 4. The length of the MD models is up to 200 nm, which, although about half the size of the experimental device, still captures its essential physical properties. To produce a stationary heat current ($J$), the ends of the device are kept at 350 and 250 K (see Methods), respectively, by two Langevin thermostats. If the system displayed thermal rectification its thermal conductance, computed as $G = J/\Delta T$, would differ if the heat current went from 1L→2L or from 2L→1L. Setting up the NEMD simulations we have two options to treat the bilayer side of the GJ: we can either apply the thermostat to both layers, as in Ref. [16], or treat only the top layer as a thermal bath. In the first case we find that the thermal conductance is near that of a single graphene layer, much too large compared to the experiments (Supplementary Section 8). Thus, we focus our analysis on the second case. In fact, NEMD simulations show the thermal conductance of the device is the same, within the statistical uncertainty, regardless of the direction of the heat current. Hence, our simulations also confirm that this system does not display thermal rectification.

An analysis of the temperature profile at stationary conditions (Figure 4) shows that the top and bottom layers of the junction are thermally decoupled, and the main source of TBR is not the step at the junction, but rather the weak coupling between the two stacked graphene layers. Such weak coupling causes a larger temperature difference ($\Delta T \sim 70$ K) between the top and bottom layer of the device, whereas the temperature discontinuity at the step of the junction is only ~3 K. Hence the main resistive process occurs at the interface between the overlapping layers, which is symmetric, thus explaining why no thermal asymmetry or rectification occurs. Even with a very large temperature difference at the two ends of the device ($\Delta T \sim 450$ K), thermal rectification remains negligible (Supplementary Section 8).

Our experiments and simulations appear at odds with the NEMD results of Zhong *et al.* [15]. In this work, the system is set up such that there is no thermal decoupling between layers in the thermal reservoir, and this effect is not probed in the non-thermostated junction. Hence these former simulations suggest an asymmetric phonon scattering at the junction that depends on the heat flow direction (thermal rectification effect). By comparing our simulations with theirs, we conclude that an apparent thermal rectification could be observed by sampling the system at non-stationary conditions, stemming from poor equilibration of the thermal baths. This is especially a problem for poorly ergodic systems such as graphene and carbon nanotubes [31].

While NEMD sheds light on the microscopic details of heat transport at the GJ, it does not allow a quantitative estimate of the conductance that can be compared to experiments. In fact, due to the classical nature of MD simulations, quantum effects are not taken into account. Considering that the Debye temperature of graphene exceeds 2000 K and experiments are carried out at room temperature and below, quantum effects are expected to play a major role in determining the conductance. Thus, we also calculated the thermal conductance of the 1L-2L junction, treated as an open system, using the elastic scattering kernel method (ESKM)[32]. ESKM is an LD approach equivalent to Green's functions [33], implemented in a scalable code that allows us to compute coherent phonon transport in systems of up to $10^6$ atoms [34]. Thus, we could calculate the thermal conductance of suspended and SiO$_2$-supported GJs with the same overlap length as in the experiments. LD calculations give the phonon transmission function $\mathcal{T}(\omega)$ for an open system with semi-infinite thermal reservoirs, resolved by mode frequency and polarization. The thermal conductance is then computed by the Landauer formula [35], integrating $\mathcal{T}(\omega)$ over all frequencies:

$$G = \frac{1}{2\pi}\int_0^{\omega_{max}} d\omega\ \hbar\omega \mathcal{T}(\omega) \frac{\partial f_{BE}(\omega,T)}{\partial T}, \quad [1]$$

where *T* is the temperature and $f_{BE}$ is the Bose-Einstein distribution function, accounting for the quantum population of phonons. In this approach we neglect anharmonic phonon-phonon scattering. This assumption is justified *a posteriori* by comparing the conductance of a suspended device with overlap length of 25 nm, computed by NEMD, $G = 1.16 \pm 0.09 \times 10^9$ Wm$^{-2}$K$^{-1}$, with that obtained by LD using a classical phonon distribution function, $G = 0.92 \times 10^9$ Wm$^{-2}$K$^{-1}$. A ~20% difference between LD and NEMD calculations of $G$ is acceptable, as it may stem not only from neglecting anharmonic scattering in LD, but also from the finite Δ$T$ in NEMD.

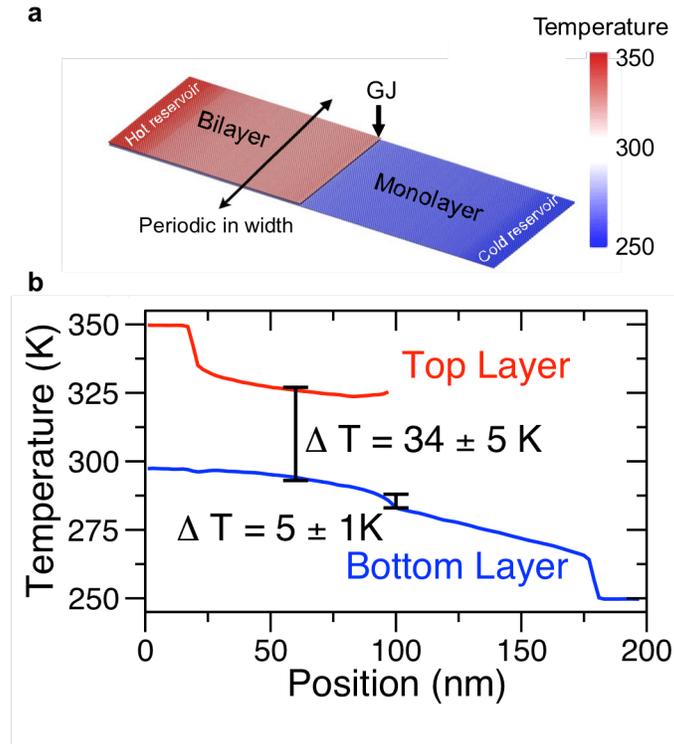

**Figure 4**. Representation of the molecular model of a suspended 1L-2L graphene junction. Atoms are color-coded according to the temperature at stationary non-equilibrium conditions. The graph shows the temperature profile in the non-equilibrium MD simulation in which the bilayer is heated to 350 K and the monolayer is cooled to 250 K. The two layers are thermally decoupled, with a major temperature difference (Δ$T$ ~ 34 K) between them; a much smaller

Figure 5 displays the thermal conductance of the 1L-2L graphene junction calculated by LD as a function of the length of the bilayer part (a) and of the temperature (b), compared to experimental data. To assess the effect of the substrate in the experimental device, we consider models of GJ both suspended and supported on a SiO$_2$ substrate. The geometry of the suspended model is the same as the one used in NEMD (Figure 4). $G$ is independent of the length of the monolayer part of the device, as in this approach it conducts heat ballistically. $G$ is normalized by the width of the GJ and by a nominal thickness of the bilayer part of 0.67 nm, which is the same convention used in processing the experimental data.

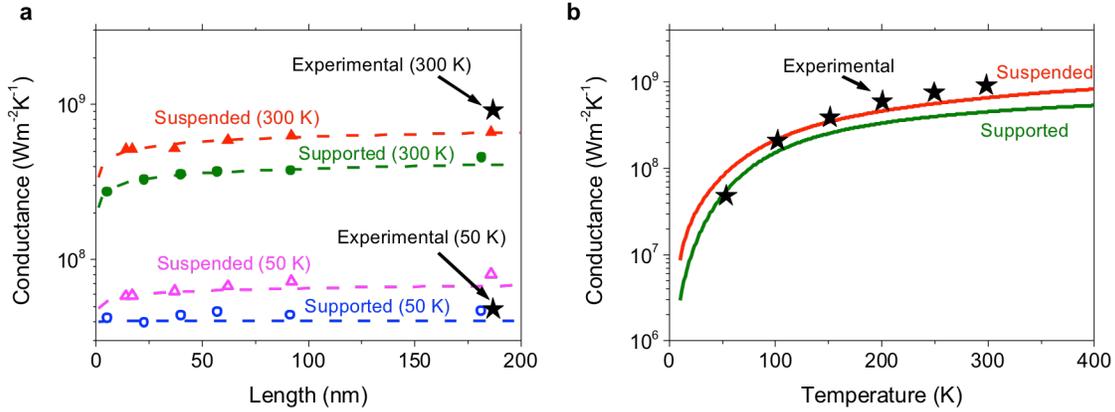

**Figure 5.** Calculated thermal conductance per unit area of a 1L-2L graphene junction, either suspended or supported on $SiO_2$ substrate. (a) Calculated conductance vs. length of the bilayer portion at 300 K and 50 K temperature (symbols). Dashed lines are guide to the eye. The stars correspond to the experimental data. (b) Calculated conductance (lines) and experimental data (star symbols) vs. temperature, both for suspended and $SiO_2$-supported GJs. The calculations use the same dimensions as in the experiment (180 nm long bilayer).

The agreement between modeling and experiments is excellent at low temperature in Figure 5. In the experimental device at higher temperature, heat transfer is still mainly dictated by the TBR between the two graphene layers, but the thermal bath also affects the bottom layer in the bilayer part of the device, thus making the conductance larger than that predicted by the model. The thermal conductance of the device increases with the interlayer overlapping surface area, which is determined by the length of the bilayer part (Figure 5a). However, $G$ does not grow linearly with the overlap surface and tends to saturate with the overlapping length. The conductance limit of this device is indeed dictated by the ballistic limit of a single graphene sheet [36].

The interaction with the $SiO_2$ substrate reduces the overall conductance of the device by about 30% at room temperature. In order to achieve quantitative agreement between theory and experiments, it is important to consider the conductance reduction in the model for supported structures. The temperature dependence of the GJ thermal conductance can be almost entirely ascribed to the quantum population of the phonon modes. In fact, Figure 5b shows that theory and experiments display an excellent agreement at low temperature, while systematic deviations appear at $T > 200$ K, allowing us to pinpoint the effect of anharmonic scattering, which is not taken into account in the calculations. Resolving the transmission function by mode polarization shows that only out-of-plane modes contribute to heat transport across the interlayer junction, consistent with another recent study [37] (Supplementary Section 9). We also observe that the interaction with the substrate causes an offset of the out-of-plane modes of the bottom graphene layer with respect to those of the top layer, thus hampering the transmission function even further and reducing the conductance of the device.

4. Conclusions

In conclusion, we have experimentally measured, for the first time, the temperature dependent heat flow across GJs, i.e. 1L-2L and 2L-4L graphene junctions, supported on $SiO_2$ substrates. MD and LD simulations were used to analyze the GJ thermal transport. The simulations show that the top and bottom layers of the junction are only weakly thermally coupled, and the main source of TBR is not the step at the junction, but rather the weak coupling between the two layers in bilayer graphene. The interaction with the substrate was observed to have a significant effect to achieve good agreement between the theory and experiments. In fact, the values obtained for the experimental and theoretical thermal conductance of supported GJs showed excellent agreement at low temperature ($T < 200$ K), whose dependence can be almost entirely ascribed to the quantum population of the phonon modes. The deviations observed above 200 K, allowed us to quantify the effect of anharmonic scattering. Additionally, the thermal decoupling observed between layers suppress the possibility of thermal rectification in GJs. Our findings shed new light on thermal transport across GJs, revealing thermal decoupling between layers that is behind the large TBR observed. These results also imply that the presence of GJs in large-area (e.g. CVD-grown) graphene should *not* affect the overall thermal conductivity of the material, unlike GB defects. Thus, the thermal properties of CVD-grown graphene are not expected to be affected by the presence of small bilayer islands, because most heat will be carried in the bottom layer.

**Methods**

Experimental measurements and data analysis

Highly crystalline graphite (carbon > 99.75%) was mechanically exfoliated with Scotch[TM] tape onto $SiO_2$ (~295 nm) on Si substrate chips of ~1 × 1 cm$^2$ size. An optical microscope was first used to find large graphene junction (GJ) samples where we could perform thermal measurements (Supplementary Section 1).

Electron-beam (e-beam) lithography (with a first layer of PMMA 495 and a second layer of PMMA 950 spin-coated on the samples at 4000 rpm for 40 seconds, and baked at 180ºC for 10 minutes) was used to pattern the heater and sensor on each side of the GJ. Heater and sensor lines are ~200 nm wide and ~5 μm long. After development, an e-beam evaporator was used to deposit 40 nm of $SiO_2$ followed by 3 nm Ti and 35 nm Pd, forming the heater and sensor lines, electrically isolated from the graphene underneath (Supplementary Section 1). The separation ($L$) between heater and sensor lines for 1L-2L and 2L-4L junction samples were $L_{1L-2L}$ = 374 nm and $L_{2L-4L}$ = 395 nm, respectively (Figure 1b and 1d).

Raman spectroscopy was carried out using a Horiba LabRam instrument with a 532 nm laser and 100× objective with N.A. = 0.9, after all fabrication and other measurements were completed. The GJ region was scanned with 150 nm step size and 160 μW laser power. The laser spot diameter obtained by the knife-edge method was < 400 nm. We analyzed the spectra of several representative locations on both sides of the GJs by removing the baseline and fitting the 2D (also known as G') peak with different Lorentzians (Figure 1c and 1e). These Raman maps determined the quality of the graphene and number of layers on each side of the GJ (also see SSupplementary Section 2).

The samples were wire-bonded into chip carriers and the thermal measurements were carried out in a cryostat at $1.3 \times 10^{-6}$ mbar, at temperatures from 50 K to 300 K (Supplementary Sections 4 to 7). Scanning Electron Microscopy (SEM) was used after thermal measurements to examine the location of the heater and sensor on each side of the GJ, as well as to measure the separation and dimensions of the lines (Supplementary Section 1).

The experimental data were analyzed using finite element modeling (FEM) with COMSOL® Multiphysics (Supplementary Information Section 10), to determine the thermal conductance of the GJ and of the various layers and interfaces. These simulations were based on previous measurements on similar samples carried out by a subset of the authors.[20],[38] The uncertainty calculations are also explained in Supplementary Information Section 10.

Non-equilibrium molecular dynamics (NEMD)

All MD simulations were carried using the LAMMPS package.[39] We used the optimized Tersoff force-field[40] for the in-plane interactions, and a Lennard-Jones (LJ) potential with $\epsilon =$ 3.29567 meV and $\sigma = 3.55$ Å for the interlayer interactions, according to the OPLS-AA parameterization [41]. Equations of motion were integrated with a time step of 1 fs. The simulated structure had a periodic width of 5 nm and interlayer spacing of 0.335 nm, containing 14736 C atoms in a 25 nm long top layer (4896 atoms) and a 50 nm bottom layer (9840 atoms) in the transport direction. Boundary conditions were fixed in the transport direction and periodic in both perpendicular directions. We first equilibrated the system in the canonical ensemble at 300 K using the stochastic velocity rescaling algorithm [42] for 0.1 ns (Supplementary Section 8).

To enable a stationary heat current, the 10% C atoms at the left end of the top layer and 10% atoms at the end of the bottom layer were thermostatted to the target temperatures of 350 and 250 K, respectively, using Langevin thermostats with a 0.05 ps relaxation time. We have tested different coupling constants and verified that a weaker coupling, e.g. 1 ps, is insufficient for the thermal baths to reach the target temperatures [31]. The first two rows of C atoms in the top and bottom sheets and the last two rows of C atoms in the bottom sheet were constrained at fixed positions, and the system was allowed to run for a total of 40 ns. The temperature profile was grouped into 100 bins along the transport direction, sampled every 10$^{th}$ step, the total average was computed every 1000 steps and the temperature was calculated from the kinetic energy. The power supplied or subtracted by the hot or cold Langevin thermal baths is averaged over time at stationary conditions to give the steady-state heat flux. The temperature profiles of the converged steady-state are averaged and plotted, and the difference in bath temperature gives the total temperature differential (Supplementary Section 8).

Lattice dynamics (LD) calculations

We compute thermal boundary conductance in the quantum regime for GJs models using LD and the elastic scattering kernel method [32]. We consider both suspended and supported junctions. The interatomic potentials used for LD calculations were the same as in the NEMD simulations for the suspended device. The interatomic interactions of the quartz substrate in the supported device are modeled with the potential by van Beest *et al.* [43]. The interactions between the graphene layers and the substrate are modeled with a LJ potential with interaction cut-offs set to 8 Å. All models had a periodic width of 4.984 nm and varying lengths. The overlap lengths for the suspended graphene junctions were 14, 17, 37, 62, 92, and 186 nm. The overlap lengths for the supported graphene junctions were 5.1, 22.4, 39.7, 56.9, 91,4 and 181.3 nm (Supplementary Section 9).

**Acknowledgements**

We acknowledge the Stanford Nanofabrication Facility (SNF) and Stanford Nano Shared Facilities (SNSF) for enabling device fabrication and measurements. We acknowledge Woosung Park for his help setting the cryostat. This work was supported by the National Science Foundation Engineering Research Center for Power Optimization of Electro Thermal Systems (POETS) with cooperative agreement EEC-1449548, by the National Science Foundation (NSF) EFRI 2-DARE grant 1542883, the Air Force Office of Scientific Research (AFOSR) grant FA9550-14-1-0251, the National Research Foundation of Korea grant 2015R1A2A1A10056103, and in part by the Stanford SystemX Alliance.

**Additional information**

Supplementary Information is available in the online version of the paper.

**Competing financial interests**

The authors declare no competing financial interests.

**References:**

[1]     Houssa M, Dimoulas A and Molle A 2016 *2D Materials for Nanoelectronics* ed Taylor & Francis Group (Boca Raton: CRC Press)

[2]     Allen M J, Tung V C and Kaner R B 2010 Honeycomb Carbon : A Review of Graphene *Chem. Rev.* **110** 132–45

[3]     Pop E, Varshney V and Roy A K 2012 Thermal properties of graphene: Fundamentals and applications *MRS Bull.* **37** 1273–81

[4]     Xu Y, Li Z and Duan W 2014 Thermal and Thermoelectric Properties of Graphene *Small* **10** 2182–99

[5]     Peres N M R 2010 Colloquium : The transport properties of graphene: An introduction *Rev. Mod. Phys.* **82** 2673–700

[6]     Huang P Y, Ruiz-Vargas C S, van der Zande A M, Whitney W S, Levendorf M P, Kevek J W, Garg S, Alden J S, Hustedt C J, Zhu Y, Park J, McEuen P L and Muller D A 2011 Grains and grain boundaries in single-layer graphene atomic patchwork quilts *Nature* **469** 389–92

[7]     Tsen A W, Brown L, Levendorf M P, Ghahari F, Huang P Y, Havener R W, Ruiz-Vargas C S, Muller D A, Kim P and Park J 2012 Tailoring Electrical Transport Across Grain Boundaries in Polycrystalline Graphene *Science (80-. ).* **336** 1143–6

[8]     Yasaei P, Fathizadeh A, Hantehzadeh R, Majee A K, El-Ghandour A, Estrada D, Foster C, Aksamija Z, Khalili-Araghi F and Salehi-Khojin A 2015 Bimodal Phonon Scattering in Graphene Grain Boundaries *Nano Lett.* **15** 4532–40

[9]     Wei Y, Wu J, Yin H, Shi X, Yang R and Dresselhaus M 2012 The nature of strength enhancement and weakening by pentagon–heptagon defects in graphene *Nat. Mater.* **11** 759–63

[10]    Serov A Y, Ong Z-Y and Pop E 2013 Effect of grain boundaries on thermal transport in graphene *Appl. Phys. Lett.* **102** 033104

[11]   Fan Z, Hirvonen P, Pereira L, Ervasti M, Elder K, Donadio D, Harju A and Ala-Nissila T 2017 Bimodal grain-size scaling of thermal transport in polycrystalline graphene from large-scale molecular dynamics simulations *Nano Lett.* **17** 5919–24

[12]   Giannazzo F, Deretzis I, La Magna A, Roccaforte F and Yakimova R 2012 Electronic transport at monolayer-bilayer junctions in epitaxial graphene on SiC *Phys. Rev. B* **86** 235422

[13]   Xu X, Gabor N M, Alden J S, van der Zande A M and McEuen P L 2010 Photo-Thermoelectric Effect at a Graphene Interface Junction *Nano Lett.* **10** 562–6

[14]   Wang X, Xie W, Chen J and Xu J 2014 Homo- and Hetero- p–n Junctions Formed on Graphene Steps *ACS Appl. Mater. Interfaces* **6** 3–8

[15]   Zhong W, Huang W, Deng X and Ai B 2011 Thermal rectification in thickness-asymmetric graphene nanoribbons *Appl. Phys. Lett.* **99** 193104

[16]   Rajabpour A, Fan Z and Vaez Allaei S M 2018 Inter-layer and intra-layer heat transfer in bilayer/monolayer graphene van der Waals heterostructure: Is there a Kapitza resistance analogous? *Appl. Phys. Lett.* **112** 233104

[17]   Ma T, Liu Z, Wen J, Gao Y, Ren X, Chen H, Jin C, Ma X, Xu N, Cheng H and Ren W 2017 Tailoring the thermal and electrical transport properties of graphene films by grain size engineering *Nat. Commun.* **8** 14486

[18]   Pop E 2010 Energy dissipation and transport in nanoscale devices *Nano Res.* **3** 147–69

[19]   Islam S, Li Z, Dorgan V E, Bae M-H and Pop E 2013 Role of Joule Heating on Current Saturation and Transient Behavior of Graphene Transistors *IEEE Electron Device Lett.* **34** 166–8

[20]   Bae M, Li Z, Aksamija Z, Martin P N, Xiong F, Ong Z, Knezevic I and Pop E 2013 Ballistic to diffusive crossover of heat flow in graphene ribbons *Nat. Commun.* **4** 1734

[21]   Jang W, Chen Z, Bao W, Lau C N and Dames C 2010 Thickness-Dependent Thermal Conductivity of Encased Graphene and Ultrathin Graphite *Nano Lett.* **10** 3909–13

[22]  Li Z, Bae M-H and Pop E 2014 Substrate-supported thermometry platform for nanomaterials like graphene, nanotubes, and nanowires *Appl. Phys. Lett.* **105** 023107

[23]  Asheghi M, Kurabayashi K, Kasnavi R and Goodson K E 2002 Thermal conduction in doped single-crystal silicon films *J. Appl. Phys.* **91** 5079–88

[24]  Chen Z, Jang W, Bao W, Lau C N and Dames C 2009 Thermal contact resistance between graphene and silicon dioxide *Appl. Phys. Lett.* **95** 161910

[25]  Koh Y K, Bae M-H, Cahill D G and Pop E 2010 Heat Conduction across Monolayer and Few-Layer Graphenes *Nano Lett.* **10** 4363–8

[26]  Cahill D G 1990 Thermal conductivity measurement from 30 to 750 K: the 3ω method *Rev. Sci. Instrum.* **61** 802–8

[27]  Yamane T, Nagai N, Katayama S and Todoki M 2002 Measurement of thermal conductivity of silicon dioxide thin films using a 3ω method *J. Appl. Phys.* **91** 9772

[28]  Seol J H, Jo I, Moore A L, Lindsay L, Aitken Z H, Pettes M T, Li X, Yao Z, Huang R, Broido D, Mingo N, Ruoff R S and Shi L 2010 Two-Dimensional Phonon Transport in Supported Graphene *Science (80-. ).* **328** 213–6

[29]  Sadeghi M M, Jo I and Shi L 2013 Phonon-interface scattering in multilayer graphene on an amorphous support *Proc. Natl. Acad. Sci.* **110** 16321–6

[30]  Jund P and Jullien R 1999 Molecular-dynamics calculation of the thermal conductivity of vitreous silica *Phys. Rev. B* **59** 13707–11

[31]  Chen J, Zhang G and Li B 2010 Molecular Dynamics Simulations of Heat Conduction in Nanostructures: Effect of Heat Bath *J. Phys. Soc. Japan* **79** 074604

[32]  Duchemin I and Donadio D 2011 Atomistic calculation of the thermal conductance of large scale bulk-nanowire junctions *Phys. Rev. B* **84** 115423

[33]  Young D A and Maris H J 1989 Lattice-dynamical calculation of the Kapitza resistance between fcc lattices *Phys. Rev. B* **40** 3685–93

[34]    Duchemin I and Donadio D 2012 Atomistic simulations of heat transport in real-scale silicon nanowire devices *Appl. Phys. Lett.* **100** 223107

[35]    Imry Y and Landauer R 1999 Conductance viewed as transmission *Rev. Mod. Phys.* **71** S306–12

[36]    Saito K, Nakamura J and Natori A 2007 Ballistic thermal conductance of a graphene sheet *Phys. Rev. B* **76** 115409

[37]    Ong Z-Y, Qiu B, Xu S, Ruan X and Pop E 2018 Flexural resonance mechanism of thermal transport across graphene-SiO 2 interfaces *J. Appl. Phys.* **123** 115107

[38]    Li Z, Bae M-H and Pop E 2014 Substrate-supported thermometry platform for nanomaterials like graphene, nanotubes, and nanowires *Appl. Phys. Lett.* **105** 023107

[39]    Plimpton S 1995 Fast Parallel Algorithms for Short-Range Molecular Dynamics *J. Comput. Phys.* **117** 1–19

[40]    Lindsay L and Broido D A 2010 Optimized Tersoff and Brenner empirical potential parameters for lattice dynamics and phonon thermal transport in carbon nanotubes and graphene *Phys. Rev. B - Condens. Matter Mater. Phys.* **81** 1–6

[41]    Robertson M J, Tirado-Rives J and Jorgensen W L 2015 Improved Peptide and Protein Torsional Energetics with the OPLS-AA Force Field *J. Chem. Theory Comput.* **11** 3499–509

[42]    Bussi G, Donadio D and Parrinello M 2007 Canonical sampling through velocity rescaling *J. Chem. Phys.* **126** 014101

[43]    van Beest B W H, Kramer G J and van Santen R A 1990 Force fields for silicas and aluminophosphates based on ab initio calculations *Phys. Rev. Lett.* **64** 1955–8